\documentclass[10pt]{article}
\usepackage[margin=1in]{geometry}
\usepackage[utf8]{inputenc}
\usepackage{natbib}
\usepackage{amsmath}
\usepackage{amssymb}
\usepackage{hyperref}
\usepackage{verbatim}

\usepackage{xcolor}
\definecolor {darkred}{rgb}{0.55,0,0}
\definecolor {darkorange}{rgb}{1,0.55,0}
\definecolor{compblue}{rgb}{0,0.45,1}
\definecolor{darkblue}{rgb}{0,0,0.55}
\definecolor{violet}{rgb}{0.55,0,1}

\usepackage{gensymb}
\usepackage{textcomp}

\usepackage[framemethod=TikZ]{mdframed}
\usepackage{newfloat}

\mdfsetup{skipabove=\topskip,skipbelow=\topskip,nobreak=false}

\tikzstyle{titregris} =
    [draw=gray!40, thick, fill=gray!10,%
        text=black, rectangle,  
        right,minimum height=.7cm]

\pgfdeclarehorizontalshading{exercisebackground}{100bp}{color(0bp)=(green!40);color(100bp)=(black!5)}

\makeatletter
\def\mdf@@exercisepoints{}
\define@key{mdf}{exercisepoints}{
    \def\mdf@exercisepoints{#1}
}
\def\mdf@@myboxedtitle{}
\define@key{mdf}{myboxedtitle}{
    \def\mdf@myboxedtitle{#1}
}
\mdfdefinestyle{boxstyle}{%
    outerlinewidth=1em,%
    outerlinecolor=white,%
    leftmargin=-1em,%
    rightmargin=-1em,%
    middlelinewidth=1.2pt,%
    roundcorner=5pt,%
    linecolor=gray!70,%
    apptotikzsetting={\tikzset{mdfbackground/.append style={gray!10}}},
    innertopmargin=1.2\baselineskip,
    skipabove={\dimexpr0.5\baselineskip+\topskip\relax},
    skipbelow={-1em},
    needspace=3\baselineskip,
    frametitlefont=\sffamily\bfseries,
    settings={\global\stepcounter{box}},
    singleextra={%
        \node[titregris,xshift=1cm] at (P-|O) %
            {~\mdf@frametitlefont{\mdf@myboxedtitle}~};
},%
    firstextra={%
\node[titregris,xshift=1cm] at (P-|O) %
{~\mdf@frametitlefont{\mdf@myboxedtitle}~};
},
}
\makeatother

\newcounter{mybox}
\newcommand{\thebox}{\arabic{mybox}}

\usepackage[capitalize,noabbrev]{cleveref}
\crefname{\thesection}{section}{sections}
\crefname{box}{Box}{Boxes}

\usepackage{authblk}
\title{Source Invariance and Probabilistic Transfer:\\A Testable Theory of Probabilistic Neural Representations}
\author[1,6,*]{Samuel Lippl}
\author[1,7,*]{Raphael Gerraty}
\author[1,2,8,**]{John Morrison}
\author[1,3,4,5,9,**]{Nikolaus Kriegeskorte}
\affil[1]{Zuckerman Mind Brain Behavior Institute, Columbia University}
\affil[2]{Department of Philosophy, Barnard College, Columbia University}
\affil[3]{Department of Psychology, Columbia University}
\affil[4]{Department of Neuroscience, Columbia University}
\affil[5]{Department of Electrical Engineering, Columbia University}
\affil[6]{Email: \texttt{samuel.lippl@columbia.edu}}
\affil[7]{Email: \texttt{rgerraty@gmail.com}}
\affil[8]{Email: \texttt{jmorriso@barnard.edu}}
\affil[9]{Email: \texttt{nk2765@columbia.edu}}
\affil[*]{Joint first author}
\affil[**]{Joint senior author}

\usepackage{hyperref}       
\hypersetup{citecolor=darkred, urlcolor=compblue, linkcolor=darkblue, colorlinks=true}

\begin{document}

\newcounter{box}
\mdfsetup{innertopmargin=10pt,linecolor=grey!40,%
linewidth=2pt,topline=true,%
frametitleaboveskip=\dimexpr-\ht\strutbox\relax,nobreak=false}

\newenvironment{mybox}[2][]{%
\refstepcounter{mybox}%
\ifstrempty{#1}%
{\mdfsetup{%
frametitle={%
\tikz[baseline=(current bounding box.east),outer sep=0pt]
\node[anchor=east,rectangle,fill=blue!20]
{\strut Box~\thebox.};}}
}%

\begin{mdframed}[style=boxstyle,nobreak=false,myboxedtitle={Box~\thebox.~#1}]\relax%
\label{#2}}{\end{mdframed}}
\maketitle

\renewcommand{\abstractname}{Short abstract}
\begin{abstract}
    As animals interact with their environments, they must infer properties of their surroundings. Some animals, including humans, can represent uncertainty about those properties. But when, if ever, do they use probability distributions to represent their uncertainty? It depends on which definition we choose. In this paper, we argue that existing definitions are inadequate because they are untestable. We then propose our own definition, which defines probabilistic representations in terms of two properties: 1) invariance to the source of uncertainty and 2) consistency in how this uncertainty is taken into account by downstream computations across multiple tasks.
\end{abstract}

\renewcommand{\abstractname}{Long abstract}
\begin{abstract}
As animals interact with their environments, they must infer properties of their surroundings. Some animals, including humans, can represent uncertainty about those properties. But when, if ever, do they use probability distributions to represent their uncertainty? It depends on which definition we choose. In this paper, we argue that existing definitions are inadequate because they are untestable. We then propose our own definition.

There are two reasons why existing definitions are untestable. First, they do not distinguish between representations of uncertainty and representations of variables merely related to uncertainty (`representational indeterminacy'). Second, they do not distinguish between probabilistic representations of uncertainty and merely ``heuristic'' representations of uncertainty. We call this `model indeterminacy' because the underlying problem is that we do not have access to the animal’s generative model.

We define probabilistic representations by two properties: 1) they encode uncertainty regardless of the source of the uncertainty (`source invariance'), 2) they support the efficient learning of new tasks that would be more difficult to learn given non-probabilistic representations (`probabilistic task transfer'). Source invariance indicates that they are representations of uncertainty rather than variables merely related to uncertainty, thereby solving representational indeterminacy. Probabilistic task transfer indicates that they are probabilistic representations of uncertainty rather than merely heuristic representations, thereby solving model indeterminacy.
\end{abstract}

\newpage

\section{Introduction}
\label{sec:introduction}

You are about to cross a street when you see a car approaching. Should you cross anyway?
One approach is to estimate the time it would take you to cross the street (say, five seconds) as well as the distance and speed of the car, so as to determine if you can make it across in time. If you predict that the car will take longer to arrive than it takes you to cross, you cross; otherwise, you wait. But this approach is risky: your estimates of the car's distance and speed (as well as the time it would take you to cross the street) may be incorrect --- and a wrong decision could kill you.

Uncertainty poses a fundamental challenge for perceptual systems \citep{smith_alhacens_2001,mach_contributions_1897,helmholtz_treatise_1924}. There are different ways in which the brain could be robust to uncertainty. One solution is to be broadly aware that the perceptual system may make errors in its estimates in making decisions. So if you predict that it takes you five seconds to cross the street but it takes the car six seconds to reach your position, you may still want to avoid crossing the street.

However, our brains actually do better. Our uncertainty depends, in part, on external conditions: in the dark or under foggy weather conditions, we will generally be less certain about our estimates \citep{bach_knowing_2012} and should therefore leave more leeway for potential perceptual errors. Representing our uncertainty about the car's speed and distance would allow us to adjust our behavior to our current level of uncertainty and prevent us from being overly (or insufficiently) cautious. For instance, if we estimate that the car will arrive at our location in about ten seconds, we may be fine with taking this risk on a sunny day with clear viewing conditions, but on a rainy night, we may want to wait for the car to pass.

Some evidence that humans and other animals might represent their uncertainty comes from experiments in which they have to make a choice between two options for a reward.
For example, a subject might use an eye movement to indicate the direction in which a set of dots is moving on average.
A correct choice yields a large reward and an incorrect choice no reward. When human and non-human subjects are given the choice to “opt-out” for an intermediate reward, they choose that option when they are not confident, for example on trials with very brief duration \citep{kiani_representation_2009,zizlsperger_cortical_2014,gherman_neural_2015}.
Opting out when the certain intermediate reward beats the uncertain large reward is the correct strategy for maximizing the expected total reward in the long term.
That subjects use such a strategy suggests they can represent their uncertainty.

Given that you represent your uncertainty (about, say, the car's speed and distance), \textit{how} should you do it? A normative theory of uncertainty representations is probability theory, a formal language that is provably optimal for making inferences under uncertainty \citep{ramsey_truth_1926,kolmogorov_sulla_1933,cox_probability_1946}. Many researchers have claimed that humans and other animals ``use probability distributions'' to make decisions under uncertainty \citep{ma_neural_2014}. Indeed, representations of probability distributions (i.e.\ ``probabilistic representations'') can predict behavior in tasks like the one described above. More generally, probabilistic accounts of perception are widespread in psychology, neuroscience, and related fields, and have been successful at predicting brain activity and behavior across a wide range of domains \citep{hurlimann_testing_2002,tenenbaum_theory-based_2006,chater_probabilistic_2008,griffiths_probabilistic_2010,fiser_statistically_2010,vilares_bayesian_2011,ma_organizing_2012,oreilly_how_2012}.

However, while it is generally undisputed that humans and animals can represent their perceptual uncertainty, theories of perception positing that they do so using probability distributions have long been controversial.
Some researchers have highlighted that almost any behavior can be explained by some probabilistic model.
The freedom to pick suitable prior assumptions for a given behavior seems to render probabilistic accounts unfalsifiable \citep{jones_bayesian_2011,bowers_bayesian_2012,marcus_how_2013}.

One response is to require not only optimal inference given a set of prior assumptions, but also an optimal set of prior assumptions.
For example, when a subject estimates the speed of a car, we may require that they use the optimal prior over car speeds, i.e.\ that which represents the ground-truth distribution and leads to the best possible inference. Such an optimal-observer assumption would allow us to test for probabilistic representations. For example, in studies of perceptual cue combination, humans and primates weigh evidence from different sources by their uncertainty in a way that appears to approximate the optimal model \citep{ernst_humans_2002,gu_neural_2008}. An optimal-observer view suggests that their representations of these cues are probabilistic.
In signal detection tasks, in contrast, humans and primates often do not respond to uncertainty as dedicated by the optimal model, perhaps because they are relying on inaccurate estimates of the true frequencies of signals in the environment \citep{kahneman_prospect_1979,maddox_overestimation_1998,peters_perceptual_2017}. Likewise, humans and other organisms are often miscalibrated in their confidence, that is they systematically over- or underestimate the accuracy of their own decisions \citep{baranski_calibration_1994,kiani_choice_2014}. An optimal-observer view suggests that they are therefore not using probabilistic representations. They are instead using ``heuristic'' representations of uncertainty \citep{gigerenzer_why_2008}.

We’ll argue that this response is unsatisfactory. To start, it is often unclear what it means for a probabilistic representation to be optimal, especially outside the context of a specific experimental task. Moreover, it is unclear why optimality should distinguish between heuristic and probabilistic representations of uncertainty. For example, probabilistic models in statistical analysis often rely on a wrong generative model, yet no one would consider them not probabilistic for this reason.

More generally, the present debate on probabilistic representations faces a key dilemma: If we don't introduce more specific constraints, we can describe \textit{any} behavioral and neural data in terms of probabilistic representations. We refer to this as the issue of \textbf{model indeterminacy}. The constraints that have been suggested, however, are not well-motivated by probability theory and do not provide a clear distinction between probabilistic representations and heuristic representations of uncertainty.

Given the complications of defining probabilistic representations as an empirically falsifiable proposition, it is tempting to regard the debate about them as purely semantic.
Perhaps probability theory serves merely as a \textit{language} to describe representations of uncertainty rather than as a source of a testable hypothesis about what the brain is representing.
Indeed, the existing literature on probabilistic representations can be read through such a lens. The evidence for and against probabilistic representations contributes to an overall picture that helps us better understand how the brain takes into account its uncertainty about the world. Probabilistic inference certainly provides a useful normative reference for understanding brain computation. However, there is a reason probability theory has a special status among the strategies that enable an agent to take uncertainty into account. The separation of knowledge (albeit uncertain) and action enables generalization of internal representations of the world to novel stimuli and task challenges.

In this article we develop an empirically falsifiable theory of probabilistic representation that is rooted in the separation of knowledge and action at the heart of probabilistic inference and probabilistic decision theory, which enables internal representations to generalize across both sources of sensory uncertainty and target tasks. By focusing on the essential power of probabilistic inference to generalize across stimuli and responses, our proposal manages to avoid the pitfalls discussed above \citep[see also][]{maloney_bayesian_2009,koblinger_representations_2021}.
\cref{sec:defining} surveys existing attempts to define probabilistic perceptual representations and shows that none of them are able to overcome the issue of model indeterminacy. It further highlights a secondary challenge concerning the investigation of the neural implementation of representations of uncertainty, the challenge of \textbf{representational indeterminacy}.
\cref{sec:our-theory} introduces our theory which defines probabilistic representations in terms of two principles: \textbf{source invariance} and \textbf{probabilistic transfer}. We argue that these principles can overcome these indeterminacies.
\cref{sec:evidence} concludes the article with a brief survey of questions that arise from our definition; existing findings in the fields of psychology, neuroscience, and machine learning that can speak to them; and a broad range of remaining open questions.

\section{Defining Probabilistic Representation}
\label{sec:defining}

Probability theory is a normative theory for reasoning under uncertainty. Its axioms precisely define probability distributions. But they do not define what it is to represent a probability distribution. 

An adequate definition must answer three questions. First, what is it to represent a probability distribution, as opposed to a non-probabilistic value? Most researchers agree that it is possible to represent a probability distribution by representing some aspect of that distribution. But which aspects? Depending on our definition, a probabilistic representation might, for example, represent the entire distribution, random samples from the distribution (with the approximation becoming more precise with more samples), parameters such as mean and variance, or even just the point estimate with the most likely value.
 
Second, what counts as representing a probability distribution, as opposed to a non-representational relation to probability distributions? As we’ll see, all activity can be described in terms of probability theory. But we want a definition that lets us distinguish between the activity that represents a probability distribution and activity that can merely be described in terms of probability theory. 
 
Third, what constraints (if any) are there on the generative models underlying probability distributions? Most researchers agree that a probabilistic representation is, in some way, computed on the basis of and used in conjunction with an underlying ``generative model''. For example, we cannot observe the car's arrival time in our example directly, so in order to represent a feature of a probability distribution over arrival times, we need a way to infer the feature from some observable inputs, such as the activity of neurons in our retina. Probability theory tells us how to do this, but it requires formalizing our assumptions about the problem using a generative model: a joint probability distribution over the variables we can directly observe (such as retinal activation in this example) and the variables we would like to infer (the car's arrival time in this case). Given a generative model, probability theory tells us how to compute the represented aspects of the probability distribution over the represented feature and how to use these aspects in downstream computations. This process is called probabilistic inference.  Some definitions of probabilistic representation place constraints on which generative models can be used in probabilistic inference.

As  we’ll see, different definitions of probabilistic representations provide different answers to these three questions. 

\subsection{Representations are probabilistic by virtue of their role in computation}
\label{sec:role-in-computation}

\textit{What does it mean to specify a probability distribution?}

\begin{figure}
    \centering
    \includegraphics{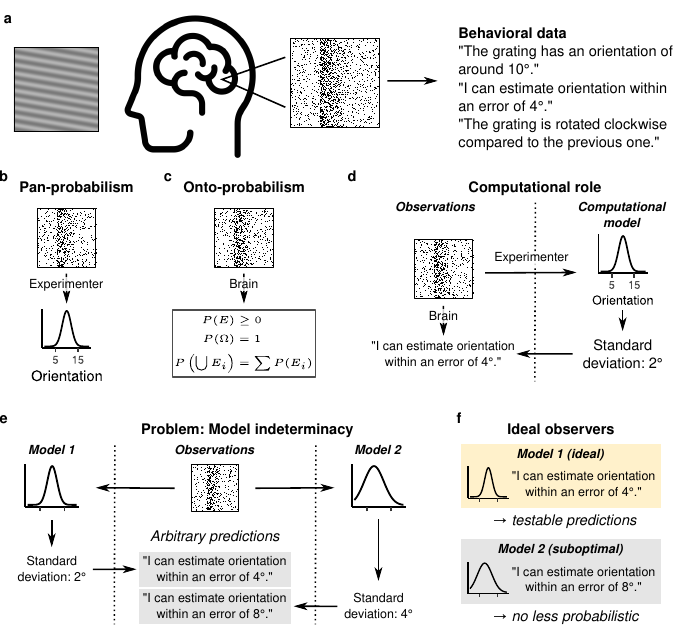}
    \caption{Defining probabilistic representation. \textbf{a}, The basic setup: Subjects perceive a certain input, which may be uncertain. For example, they may perceive a grating with a certain orientation. Their neural representation (represented schematically by a spikeplot) produces different behavior. Is this representation probabilistic? \textbf{b}, Pan-probabilism postulates that since any representation can be used as evidence in probabilistic inference, all representations are probabilistic. \textbf{c}, We may require that probabilistic representations must use the abstract mathematical machinery that defines probability distributions. This would exclude most uses of probabilistic representations in humans, animals, and neural networks. \textbf{d}, Defining probabilistic representations by virtue of their role in computation suggests that we can understand their \emph{computations} through the lens of probability. \textbf{e}, This definition still struggles with pan-probabilism: we can always come up with a generative model that can explain a given behavior. \textbf{f}, Ideal observer models provide one way of avoiding model uncertainty: by choosing the optimal generative model. However, any non-optimal generative model is no less probabilistic.}
    \label{fig:defining}
\end{figure}

\noindent Any representation of a variable can be used to define a probability distribution over that variable.
Consider, for example, a single neuron that represents the car’s speed by firing at a rate proportional to the speed.
If we postulate that the neuron encodes just a point estimate, we will not consider it a probabilistic representation.
However, an ideal observer downstream could also use the firing rate as ``evidence'' to infer a probability distribution over speed, e.g. by combining the firing rate with assumptions about the noise affecting it and a prior distribution.
This leads us into the trap of \emph{pan-probabilism}, where since any representation can be treated as evidence, any representation is probabilistic (\cref{fig:defining}b).
Put differently, this view collapses the concept of probabilistic representations into the concept of representations in general.
When describing how the neurons' firing rates can be used in downstream areas, it can be helpful to describe representations in this way \citep{oram_ideal_1998,chen_overview_2013}.
But an empirically testable definition of probabilistic representation will have to be more constrained.

Perhaps this constraint can come from the nature of a probability distribution.
Probability distributions are abstract objects defined by a set of sophisticated mathematical tools.
When we talk of probability distributions or use them in our statistical models, we use this set of sophisticated tools.
One way to conceive of the brain representing a probability distribution would be in terms of the same tools (\cref{fig:defining}c).
But this would require, at minimum, representing the basic mathematical rules underlying probability theory and the mathematical machinery used to define probability distributions.
Defining probabilistic representation this way would capture humans explicitly thinking about probability theory, but would likely exclude humans or animals involved in other tasks.
This is not the distinction we are looking for in a theory of probabilistic representation that extends to animals, individual neural populations, and even artificial neural networks.
Crucially, the apparent dichotomy between a restricted definition that appeals to idealized mathematical objects and an unrestricted definition that entails pan-probabilism is one source of skepticism about probabilistic representations.

But there is another way to think about the problem.
Probability theory does not just define probability distributions, but also a set of rules to construct and use these distributions.
Under a given generative model, these rules tell us how to construct probability distributions from our observations, and how to use them in downstream computations.
Perhaps, being a probabilistic representation is a matter of approximating the mapping from observations to an aspect of the probability distribution, or the mapping from an aspect of the probability distribution to downstream representations and behaviors \cite[see also][]{shea_neural_2014,lee_representing_2022}. This builds on a popular view on representations in general called ``conceptual role semantics" \citep{harman_conceptual_1982,harman_nonsolipsistic_1987,block_functional_1988}.
Under this view the generative model and the general rules of inference do not need to be represented, clearing the concerns that arose in the previous paragraph. As long as the computations approximate the rules of probabilistic inference in the context of the specific generative model, the representation is probabilistic (\cref{fig:defining}d).
Although rarely stated explicitly, this is the view most experimental investigations of probabilistic representations seem to take.

This approach also raises a fundamental challenge, however: if the generative model does not need to be explicitly represented, we need to additionally infer this generative model to ascertain whether a representation follows the rules of probability.
As we will argue in detail below, given any computation, we can always find a generative model such that the computation implements probabilistic inference (\cref{fig:defining}e).
We call this issue \emph{model indeterminacy}, and it seems to imply that all representations are probabilistic.
Thus pan-probabilism remains a worry for use-based definitions.

\subsection{MAP estimation and priors cannot be distinguished from optimizing an objective function}
\label{sec:map}

\textit{What aspects of a probability distribution must be represented?}

\noindent
The simplest version of representing an aspect of a probability distribution is to represent its most likely value.
Computing a variable’s most likely value given a generative model and a set of observations is called maximum a posteriori (MAP) estimation.
Since MAP estimates reflect single values rather than ranges, they are not always seen as probabilistic representations.
In fact, our own theory will exclude them for this reason.
Still, computations that implement MAP estimation seem probabilistic.
After all, they begin with assumptions stated in probabilistic terms and output the estimate most probable according to probability theory.

Many probabilistic models in neuroscience and psychology take the form of MAP estimation \citep{deneve_efficient_2001,ernst_humans_2002,lee_hierarchical_2003,friston_theory_2005,ma_organizing_2012}.
It is particularly appealing for ``top-down'' generative models, which describe our sensory inputs as generated by the features we would like to represent.
We can often divide this model into two parts: first, we express our assumptions about the feature’s prior distribution, $p(z)$ (where $z$ is the feature).
For example, we may assume that cars on this road drive around 50 mph and express that by a prior normal distribution centered around that value.
Second, we express our assumptions about the observation's likelihood, its conditional probability $p(x|z)$, for a given value of the feature.
From these two quantities, Bayes' theorem tells us how to infer the conditional distribution $p(z|x)$ of the feature given our observation, which we can use to determine its most probable value, the MAP estimate.

Cue combination is a prominent example.
Our generative model may reflect the assumption that the same feature in our environment causes multiple observations.
For example, when cars are closer, they are usually louder.
The car therefore causes both a visual and an auditory cue and we would hope to optimally combine these cues in estimating the car’s distance.
MAP estimation is one way of combining cues.
Although the combined estimate is a single value, that value is the mode of the posterior distribution entailed by the observation, generative model, and rules of probability.
So it is appealing to describe the combined estimate as representing an aspect of this posterior.
For this reason, representations involved in cue combination have often been called probabilistic, and evidence that MAP estimation can describe how humans and animals integrate multisensory cues is considered evidence for probabilistic representations \citep{ernst_humans_2002,seilheimer_models_2014}.
These models of cue combination have also been used to explain neural activity in brain areas that are presumed to be involved in this process \citep{fetsch_neural_2012, bizley_where_2016}.

In the case of cue combination (and some other top-down generative models) the features generating our observations constitute actual causes (such as objects and their properties) in our environment.
Other times, the features involved in our generative model are useful mathematical constructs, for example in Gaussian scale mixtures \citep{wainwright_scale_2000} and hierarchical Bayesian inference \citep{lee_hierarchical_2003, friston_theory_2005}.
MAP estimates derived from these generative models have been used to predict neural activity in visual cortex \citep{rao_predictive_1999, orban_neural_2016}.

In considering the second point of contention (\emph{Which aspects of the probability distribution must be represented?}), a natural definition of probabilistic representation might thus include the output of any computation that implements MAP estimation.
However, almost any computation implements MAP estimation for a suitably chosen model.
This leads to the unappealing result that many computations that do not seem probabilistic in fact output probabilistic representations.
Suppose, for example, you estimate the car's speed by averaging ten measurements.
This seems like an example of a non-probabilistic computation that outputs a non-probabilistic representation.
But we could also describe this average as the MAP estimate over speed for a generative model that assumes your measurements have a Gaussian likelihood function.

A more general way of looking at this issue is to consider representations that optimize an \emph{objective function}, a quantity that a system is adapted to maximize.
This has been an influential idea in all fields interested in cognition, ranging from work on efficient coding in V1 \citep{barlow_possible_1961} to the more recent idea that representations in the visual hierarchy can be explained by deep neural networks optimized for object recognition \citep{yamins_performance-optimized_2014,khaligh-razavi_deep_2014,guclu_deep_2015}.
These representations are usually not seen as probabilistic.
Yet almost any objective or loss function guiding such an optimization is proportional to posterior probability under a corresponding generative model, and minimizing that loss would be maximizing posterior probability\footnote{More specifically, suppose you learn a parameterized prediction $\hat{y}=f_{\theta}(x)$ by minimizing the loss function $l(y,\hat{y})$. This is equivalent to maximizing the probability density $p(y|\hat{y})\propto\exp(-l(y,\hat{y}))$, which would turn this model into a probabilistic computation according to the proposed definition above. The only necessary criterion for $p$ to be well-defined is that $\int_y\exp(-l(y|\hat{y}))<\infty$, which is the case for all losses that grow faster than $\log x$. This includes, for instance, the squared loss, the absolute loss, and the cross-entropy loss.\label{fn:loss}}.
Due to model indeterminacy, there is a real sense in which every computation that optimizes \emph{something} is probabilistic because it maximizes posterior probability. This view cannot provide a testable definition of probabilistic representations.

So far, we have not mentioned priors.
It is tempting to think that  this will help deal with this ambiguity.
Indeed, taking into account prior knowledge has sometimes been suggested as the defining feature of Bayesian inference \citep{sohn_bayesian_2019}.
But computations that rely on prior knowledge are still subject to model indeterminacy.
Suppose you know that cars typically drive down the road at 20 mph.
When you measure a car's speed, you might use this information by taking a weighted average of the 20 mph and your measurement.
Again, this does not seem like a probabilistic computation.
Yet it implements MAP estimation if we assume a prior over a car's speed that is normally distributed around 20 mph, and a normally distributed measurement error around the car’s actual speed.

More generally, using a prior to inform our current estimate is essentially imposing an inductive bias.
Imposing an inductive bias does not, by itself, require invoking probability theory.
For example, we could also imagine correcting our estimated speed towards the speed we know cars usually drive at on this road, without ever introducing normal distributions, likelihood functions, and Bayes' theorem.
And because of model indeterminacy, there will always be a set of assumptions that renders any inductive bias probabilistic.
Therefore there is no obvious way to constrain the set of possible inductive biases to a genuinely probabilistic subset.

Even more generally, any heuristic correction implements a probabilistic computation for a suitably chosen prior.
For example, we often impose regularization penalties on parameters we try to estimate so as to encourage smaller parameters \citep[$L_2$-penalization, the sum of squares;][]{hoerl_ridge_1970} or smaller and more sparse parameters \citep[$L_1$-penalization, the sum of absolute values;][]{santosa_linear_1986,tibshirani_regression_1996}.
Like prior distributions, these penalties depend on our assumptions about the world.
Any regularization penalty can (under mild constraints) be expressed as a prior probability distribution, as is shown in \cref{fn:loss}.
Including an $L_2$-norm (the sum of parameters' squared values) in an objective function is equivalent to MAP estimation with a Gaussian prior, while including an $L_1$-norm (to enforce sparsity using the sum of parameters' absolute values) is equivalent to performing MAP estimation with a Laplace prior distribution.
Thus, if implementing MAP is enough to be probabilistic, all of these computations are probabilistic, even though they can easily be described and justified without reference to probability.

These issues will arise even if we are able to follow the step-by-step computations that implement the potentially probabilistic model.
Any population representing the output of an objective function could also be cast as representing a posterior probability.
Likewise, it will always be ambiguous whether a part of the algorithm computes the gradient of our objective function or the gradient of our posterior probability.
So even if we were able to decode these probabilities or their gradients during inference, the problem would persist.

Some proponents of the Bayesian Brain Hypothesis (and related frameworks) may argue that this is precisely why we should think of the brain as probabilistic \citep{friston_theory_2005,friston_history_2012,clark_whatever_2013}.
According to this view, objective functions are fundamentally always about probability (usually referred to as surprise) and so, any representation optimized for an objective should be seen as probabilistic.
While this is a consistent perspective on probabilistic representations that can be useful for modeling the brain, it cannot be used to make predictions that are not also explainable from non-probabilistic perspectives \citep{orlandi_innocent_2014,block_border_2023}.

\subsection{Probabilistic representations of uncertainty, by themselves, cannot be distinguished from heuristic representations of uncertainty}
\label{sec:uncertainty}

Despite probability being the language of uncertainty, the examples of MAP estimation considered above did not actually involve representations of uncertainty.
There are two ways in which uncertainty representations could be involved in probabilistic representations.

First, we might require that an uncertainty representation be involved in \emph{computing} the probabilistic representation.
Different factors in our environment should make us more or less certain about our estimates.
For example, we may be less certain about a car's speed when it is rainy; our generative model would formalize this assumption by stating that our measurement has higher variance when it is rainy.
This would be reflected even in MAP estimation.
Indeed, as detailed in the previous section, experiments on cue combination have revealed that subjects are able to represent their uncertainty about each cue to determine their overall estimate.
Subjects will also rely on their prior distribution to a larger extent when they are more uncertain about their estimate, as probability theory would suggest \citep{gregory_eye_1973,stocker_noise_2006,jazayeri_temporal_2010}.

Second, we might define the probabilistic representation itself to be a representation of uncertainty that arises from a probabilistic inference.
Indeed, different studies have found evidence of such uncertainty representations:
In tasks where monkeys learned to opt out of a decision-making task if they were less certain about the correct decision, neurons in parietal cortex were found to represent this uncertainty \citep{kiani_representation_2009}. More broadly, subjects represent their confidence about a variety of different decisions and representations \citep{kepecs_neural_2008,vilares_differential_2012,rademaker_introspective_2012,yoo_uncertainty_2021}. Further, \citet{van_bergen_sensory_2015} were able to decode sensory uncertainty from visual cortex and demonstrated that human observers appeared to take this uncertainty into account in a probabilistic manner.
Finally, \citet{walker_neural_2020} were able to decode an estimate of orientation uncertainty from monkey V1 that was predictive of behavior.

So there is evidence that humans maintain probabilistic representations of uncertainty \citep{sanders_signatures_2016}.
Further, given the fundamental role of uncertainty in probability theory, this definition seems like a promising way of distinguishing probabilistic representations.
Indeed, our own theory will require that probabilistic representations be representations of uncertainty.
On its own, however, there are two problems with this definition.

\begin{figure}
    \centering
    \includegraphics{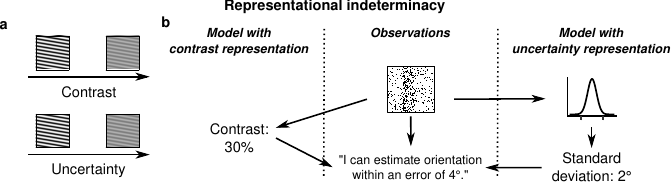}
    \caption{The issue of representational indeterminacy. \textbf{a}, A representation of uncertainty could equally be a representation of a given nuisance variable (e.g.\ contrast). \textbf{b}, Both of these models can provide an explanation for a particular representation.}
    \label{fig:rep-indet}
\end{figure}

First, it is not clear from this evidence whether humans and animals actually learn to represent uncertainty rather than an uncertainty-inducing feature of the environment.
For example, suppose you decide not to trust your estimate of when the car will arrive because it is dark outside.
We could describe you as being uncertain about the car's speed.
Or we could describe you as having learned to discount your estimate of a car's speed when it is dark outside.
Similarly, we would be able to decode a posterior distribution over speed from a representation of speed and darkness rather than speed and uncertainty about speed.
For this reason, it can be hard to determine whether a subject or brain area is representing uncertainty rather than the thing that seems to cause the uncertainty (\cref{fig:rep-indet}a).
We call this the problem of \emph{representational indeterminacy}.

This issue becomes especially important when asking whether a particular brain region is representing uncertainty.
Even if we can decode uncertainty experimentally from that region, it is possible that what is being represented is a particular nuisance variable which is used by downstream areas to compute uncertainty.
In experiments manipulating contrast to test probabilistic representations of orientation, an experimenter would be able to decode probability distributions from neural representations of estimated contrast (\cref{fig:rep-indet}b).
Indeed, this is a common experimental set-up in studies investigating probabilistic representations.

Some researchers have attempted to resolve the issue by appealing to internally generated uncertainty.
For example, \citet{walker_neural_2020} considered a set of trials where contrast (i.e. the external source of uncertainty) was fixed.
They demonstrated that activity in V1 contained information about whether the monkey judged the stimulus to have high or low uncertainty even across this set of trials.
This is consistent with an invariant representation of uncertainty, however it is also still possible the neurons represented the monkey’s estimate of contrast, which varied from trial to trial.
This estimate could still be used to compute uncertainty in some downstream region, even if it did not represent uncertainty itself \citep{morrison_third-personal_2023}.
Our theory will invoke the idea of source invariance to identify representations that are about uncertainty instead of uncertainty-inducing environmental features.

Second, even if we were able to resolve representational indeterminacy, this would still not mean that we have found a probabilistic representation.
We could also take uncertainty into account using an approximate strategy that is in some way less principled than a probabilistic computation \citep{gigerenzer_why_2008}.
Suppose, for instance, we knew that our measurement of the car's speed was uncertain and we represented this uncertainty by a positive value $a$ ($a$ being larger if we are more uncertain).
If we also knew that cars usually drive around a certain prior speed on this road, it would seem natural that our estimate is a weighted average of our measurement and our prior expectation, where the prior expectation should be weighted more strongly the larger our uncertainty (i.e. the larger $a$). For example, we could use the rule
$
\tfrac{a}{1+a}\cdot{\rm prior} + \tfrac{1}{1+a}\cdot {\rm measurement}.
$
Coming up with such heuristics would require a notion of uncertainty, but not invoke probability theory at all.
Still, for any heuristic, we could come up with a generative model that maps this heuristic onto a posterior distribution (in fact, the ``heuristic" we have just proposed would correspond to MAP estimation with a Gaussian prior with variance 1 and a Gaussian measurement error with variance $a^2$).

The same problem arises for probabilistic representations that reflect uncertainty themselves.
For example, Bayesian decision theory provides us with a probabilistic perspective on making decisions under uncertainty \citep{yuille_bayesian_1993, berger_statistical_2013}.
Decision-making usually involves a reward function $R(a,z)$ that measures the utility of a particular action $a$ in a state of the world $z$.
If $z$ were known, we could simply pick the action that maximizes reward.
For example, crossing a street would have a positive reward $r_1>0$ relative to waiting if the car did not hit us, but would have a negative reward (or punishment) $r_2<0$ if the car hit us.
So if we knew that the car would not hit us, we would decide to cross the street, and if we knew that it would, we would not cross it.

Bayesian decision theory is concerned with a situation where we are uncertain if the car will hit us when crossing the street.
For a given set of assumptions and observations, probability theory allows us to infer the probability $p$ that the car will hit us.
To choose the best action we now compute the expected reward given by $\mathbb{E}[R(a,z)]=r_1\cdot p+r_2\cdot(1-p)$.
If the expected reward of crossing the street as compared to waiting is positive, we cross the street.
This means that if we are sufficiently certain that the car will not hit us, we would decide to cross the street.
Again, we could make the same decision based on a heuristic representation of uncertainty and for any such heuristic rule, there will be a generative model that results in the same computation from a probabilistic perspective.
In fact, the complete class theorem assures us that as long as a decision rule is not suboptimal for every possible state of the world, it is consistent with posterior inference for some generative model \citep{wald_essentially_1947}.

This is all to say that any proposed definition of probabilistic representations will have to address the issue of model indeterminacy.
Most current accounts appear to, at best, circumvent this issue by defining a probabilistic representation as any representation optimizing an objective function, or any representation of uncertainty, or otherwise using probability theory as a language for understanding such representations \citep{walker_studying_2022}.
This can be a productive approach: probability theory allows us to describe any of these representations precisely due to the issue of model indeterminacy.
But such definitions cannot differentiate a probabilistic uncertainty representation from a non-probabilistic one.
At most, they are able to test whether, say, a certain neural population represents uncertainty.

More generally, it seems impossible to differentiate probabilistic from non-probabilistic representations without knowing the generative model that this probabilistic representation would instantiate.
In light of this, some researchers have proposed to constrain what makes a generative model suitable for probabilistic inference.
We will review these attempts in the next section.

\subsection{Constraints on the generative model are not grounded in the nature of probability}
\label{sec:ideal-observers}

\textit{What constraints (if any) are there on the generative models underlying probability distributions?}

\noindent
Probabilistic models have not just been a popular account of how the brain represents uncertainty.
They have also been a useful tool for researchers to characterize optimal performance on a particular task.
This is due to the fact that when faced with incomplete knowledge about task-relevant variables, probability provides the optimal way of performing a task.
So research in this field often proposes a generative model which would be optimal for performing a particular task, analyzes an upper bound on performance, and then compares this upper bound to human behavior.
These models are called ideal observers and provide useful context to a subject’s behavior \citep{geisler_ideal_2003}.

Motivated by this line of research, some have suggested that a representation is probabilistic if it supports inference under the optimal generative model \citep{mathys_bayesian_2011,feldman_bayesian_2015} (\cref{fig:defining}f).
Put differently, their response to the issue of model indeterminacy is to constrain the generative model.
Many studies leave such commitment implicit, but still argue that subjects performing as optimal Bayesian observers suggests that they have probabilistic representations \citep{ernst_humans_2002,knill_bayesian_2004}.
This strategy addresses the issue of model indeterminacy by constraining the model to be optimal.

However, the fact that probabilistic computations can be necessary for optimal behavior \citep{cox_probability_1946,ramsey_truth_1926} does not imply that optimal behavior is necessary for probabilistic computations.
Behavior might be suboptimal because a computation is relying on an inaccurate generative model.
Such a computation would still be probabilistic: probability is a language to state your assumptions and to make inferences that are consistent with them, and you can certainly use this language when those assumptions are wrong.
We use probability to analyze experimental data, for example, even though it is generally accepted that all of our models are wrong -- in fact, examining the ways in which our current probabilistic model is suboptimal is an essential part of the canonical Bayesian workflow \citep{ando_bayesian_2010}.
In a biological system, optimal probabilistic inference could lead to suboptimal behavior, under the reasonable assumption that the system's model of the world is not perfectly accurate.
To return to our car example, perhaps you have a generative model that assumes the wrong prior distribution of speeds, because you have only crossed streets inside of a city.
If you infer a posterior distribution over speed that is consistent with that model, but you are wrong because you were crossing the street on the countryside (where cars might be driving faster), it would not be fair to blame your injuries on a lack of probabilistic representation.

In fact, in most cases, a probabilistic approach is necessary \emph{because} we have suboptimal assumptions \citep{beck_not_2012}. 
For example, chess is a completely deterministic game --- there are no stochastic elements.
But, due to its complexity, an agent often has to rely on probabilistic assumptions, including assumptions about which positions are most likely to result in a win, and how one's opponent is most likely to respond to one's moves.
What often makes probability useful is precisely that we are \emph{not} ideal observers.

Another problem is that optimal behavior is impossible to define under natural conditions \citep{binmore_rational_2008,rahnev_suboptimality_2018}.
In laboratory experiments, we can create a task with a well-defined generative model and be confident that subjects have access to all information the model assumes.
This allows us to define optimal behavior.
But animals do not normally operate under such well-defined conditions.
In fact, the utility function alone can be a point of contention, with different choices resulting in substantially different predictions \citep{geisler_bayesian_2002}.

In response to these problems, it might be natural to relax the requirement that the animal's internal probabilistic model be optimal.
Perhaps it merely needs to \emph{approximate} the optimal model. But that wouldn't solve the underlying problem: It's possible to perform probabilistic inference with a generative model that's completely off. Whether the animal uses probabilistic inference must be assessed independently of whether the assumptions implicit to its behavior are optimal.

So far we have focused on the idea that an optimal model (and thus optimal behavior) is not necessary for probabilistic representation.
In fact, optimal behavior may also not be sufficient.
Any flexible enough learning algorithm can produce a function that transforms inputs into outputs that are optimal according to some decision rule.
Defining probabilistic representations in terms of optimal behavior will lead us to consider any sufficiently expressive neural network as probabilistic \citep{orhan_efficient_2017}.
More generally, when probabilistic representations are taken this broadly, their predictions will have considerable overlap with competing scientific frameworks, raising the question what unique insights they bring to the table \citep{jones_bayesian_2011}.
This will be true even for sophisticated tasks that seem to require the computation of uncertainty \citep{ma_organizing_2012}.

Attempts to constrain the generative model in different ways are problematic for the same reason.
Some accounts require testable probabilistic representations to be, in some sense, complicated, for instance by being non-normal or by having multiple modes \citep{rahnev_case_2017,rahnev_is_2020}.
Leaving aside that there is no obvious technical notion of complexity for probability distributions, probability theory as a tool is not more appropriate for expressing complex than simple probability distributions.
This is further complicated by the fact that any continuous univariate probability distribution can be expressed as a nonlinear transformation of a Gaussian random variable.

How can we reconcile these problems with the claims that probability is optimal?
The core of our framework will be defining probabilistic representations that are optimal in the same sense as probability, so this question is essential.
It turns out that probability's claims to optimality are much more humble than ideal observer models might lead us to believe.
For example, neither the Dutch Book argument \citep{ramsey_truth_1926, vineberg_dutch_2016} nor Cox's theorem \citep{cox_probability_1946,halpern_coxs_1999} (two prominent examples of optimality claims) presume a correspondence between the agent and the external world.
Rather, they show that violating the axioms of probability would result in inconsistencies, which in turn give rise to suboptimal behavior.
Importantly, as we will argue in the next section, this self-consistency is probability's distinguishing feature and it will play a central role in our definition of probabilistic representations.

\section{A Testable Theory: Source Invariance and Probabilistic Transfer}
\label{sec:our-theory}

In the previous section, we argued that standard accounts fail to distinguish probabilistic representations from non-probabilistic alternatives, identifying representational indeterminacy and model indeterminacy as crucial obstacles to such a distinction.  In this section, we will suggest 
two criteria for resolving these indeterminacies: source invariance and probabilistic transfer. Source invariance (\cref{sec:source-invariance}) resolves representational indeterminacy. A source invariant representation generalizes across different sources of the represented uncertainty (such as low contrast or high noise). By testing for source invariance, we learn whether a neural response encodes uncertainty or merely a feature correlated with uncertainty stemming from a particular source. Probabilistic transfer (\cref{sec:task-transfer}) resolves model indeterminacy. An encoding of uncertainty that enables probabilistic transfer generalizes to novel task settings. By testing for probabilistic transfer, we learn whether a neural encoding of uncertainty is used probabilistically in downstream computations. Together, source invariance and probabilistic task transfer not only enable us to test for probabilistic representations in the brain, they also enable us to locate those representations and tell us why probabilistic representations can be beneficial.

\subsection{A test for source-invariant uncertainty encoding}
\label{sec:source-invariance}

\begin{figure}
    \centering
    \includegraphics{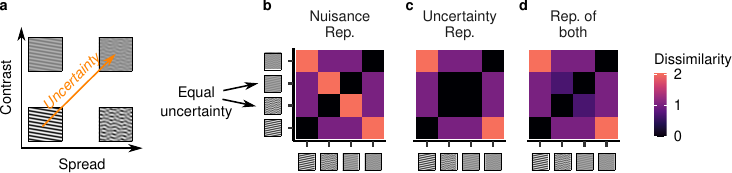}
    \caption{Source invariance. \textbf{a}, Representational indeterminacy (Figure~\ref{fig:rep-indet}) can be resolved by considering different sources of uncertainty. \textbf{b}-\textbf{d}, Representation similarity analysis provides a way of testing for source invariance. Here we consider the similarity between four different stimuli, two of which differ in their nuisance variables but have equal uncertainty. \textit{b}, A representation of nuisance variables should represent these stimuli differently. \textit{c}, An uncertainty representation should represent them similarly. \textit{d}, A population that partially represents both nuisance variables and uncertainty should lie somewhere in between these two extremes.}
    \label{fig:source}
\end{figure}

There are probabilistic representations that do not represent uncertainty (``deterministic probability distributions''). For example, a probabilistic representation of a distance of 10 meters with its entire probability mass concentrated on that value represents certainty about distance. But we cannot test for such representations because they are indistinguishable from non-probabilistic representations (in our example from a simple, non-probabilistic representation of a distance of 10 meters).  Both would lead an organism to behave as though the relevant distance were 10 meters. 

More generally, we can imagine a probabilistic representation that always represents the same level of uncertainty (other than the special case of certainty). For example, a representation of distance might always indicate a range of values $\pm$10\% of the measurement, irrespective of the sensory input. On one occasion, it might represent 9-11 meters, and on another occasion, it might represent 13.5-16.5 meters. Since the uncertainty is fully specified by the estimated value, we cannot test for whether it is represented.  They are indistinguishable from simple, non-probabilistic representations of 10 meters and 15 meters, respectively. The fixed level of uncertainty might instead be introduced by later computations.  In order to develop a testable theory of probabilistic representation, it is necessary that we restrict our focus to probabilistic representations that are capable of representing varying levels of uncertainty.

Given varying levels of uncertainty, how can we test whether a particular neural response really encodes this uncertainty? In our example of the car at night, it is hard to distinguish a representation of uncertainty about its distance from a representation of the amount of rain, because the amount of rain is a source of uncertainty and will covary with it. More generally, it is difficult to distinguish representations of uncertainty from representations of specific sources of uncertainty (\cref{sec:uncertainty}). Our proposal is to consider whether a representation is invariant across multiple sources of uncertainty, discarding information about the source of uncertainty while encoding information about the level of uncertainty (\cref{fig:source}a). In our car example, if the representation doesn’t only vary with the amount of rain, but also with the degree of darkness, and in such a way that the same representation results from multiple combinations of rain and darkness if they give rise to the same level of uncertainty, then the most parsimonious account is that it is a representation of uncertainty (\cref{fig:source}b,c). We call this ``source invariance'' and note that we are not the first to propose this criterion \citep[see e.g.][]{sahani_doubly_2003,walker_studying_2022}.

Besides being invariant to the many possible sources of uncertainty, the representation should also encode specifically how these sources affect uncertainty about a particular feature. For example, we should be able to separately read out and use uncertainty about a car's speed and its distance. This requirement, which we call ``feature specificity,'' allows us to discriminate between representations of uncertainty and general neural consequences of being in an uncertain state.

Source invariance can be experimentally tested. For example, if we want to determine whether neurons in V1 represent uncertainty about orientation, we can test whether they vary with sources of uncertainty for different combinations of contrast, spatial frequency, and darkness, among other possible sources of uncertainty. If the neurons respond equally to multiple combinations of sources that entail the same level of uncertainty and if we’re unable to decode from those neurons whether they are responding to contrast, spatial frequency, or darkness, then we can be confident that they represent uncertainty about orientation, rather than an individual nuisance variable.
Similarly, feature specificity can be experimentally tested: to make sure that those neurons represent uncertainty \emph{about orientation} as opposed to general uncertainty, we can vary uncertainty about additional properties (such as color or shape) of the stimulus and test whether the presumed uncertainty representation is invariant to that uncertainty.

Source invariance and feature specificity reflect a general principle about representation: to show that a population represents some variable (e.g. shape), we need to show that it does not represent some other variable (e.g. size) that may be correlated in our stimulus set. A comparison between responses to large squares and small circles would not be evidence that a population represents the size of an object rather than its shape. Of course, unlike size and shape, uncertainty is not an objective property of objects. Sizes and shapes would exist even without organisms, whereas uncertainty cannot. However, the same principle of invariance applies nevertheless, and indeed has been applied to other subjective perceptual representations, such as salience \citep{gottlieb_representation_1998}.

Note that source invariance is agnostic as to the \emph{neural vehicles} of the representation. For neural representations of uncertainty, it might be population vectors of firing rates, nonlinear functions of these vectors, or even another dimension of neural activity, such as temporal or spatial variability. A theory based on source invariance will be compatible with any of these specific neural vehicles.

Importantly, source invariance has benefits. If you had separate representations of rain and darkness, you would need to learn how to respond to each combination of rain and darkness for every task. If they instead produced an encoding of uncertainty, this encoding could be used across many tasks.

Source invariance is easy to define when there is no information about the source (population patterns are equal across different sources of uncertainty). However, the concept becomes more complicated once we introduce the possibility that there can be information about a source in addition to information about uncertainty.
To generalize source invariance, we may consider a representational decoding or representational similarity analyses, two frameworks that are popular in understanding representation much more broadly.
In terms of representational geometry, a neural population might cluster based on uncertainty rather than other properties (\cref{fig:source}d).
Alternately, we might say that the population represents uncertainty if we can (e.g.\ linearly) decode uncertainty more easily than the sources of uncertainty.
One way of testing this experimentally would be cross-decoding analysis.
Here, a decoder would be trained to predict a behaviorally relevant quantity that is affected by uncertainty, for example the error rate on a certain task. When tested on a new source of uncertainty (say, darkness), the decoder may make more similar predictions either for stimuli that are matched in their uncertainty (indicating a representation of uncertainty) or stimuli that are matched in their contrast (indicating a representation of contrast).
In particular, we can be confident that these analyses can address the issues introduced above to our satisfaction because they translate directly to the benefits of source invariance noted above: if an artificial decoder can generalize according to uncertainty, a downstream region reading out information from the neural population might also be more likely to generalize according to uncertainty.

Our suggested solution also implies that source invariance comes in degrees.
If uncertainty can be read out with a small amount of contamination by source information, it might be reasonable to consider the encoding invariant to the sources considered. For example, a certain neural population may reliably encode uncertainty about a car’s speed, but the decoder may still contain some information about whether this uncertainty is due to darkness or fog. The less information is left, the clearer it becomes that it’s a representation of uncertainty. Moreover, a representation may be invariant to some sources of uncertainty but not to others. For example, our representation of uncertainty about the time it will take the car to reach us may be invariant to darkness and fog but not distance and speed. How many sources does it need to be invariant to in order to be considered a representation of uncertainty? There is no definite number. The more sources, the clearer it becomes that it’s a representation of uncertainty, especially when those sources do not seem to have anything else in common.

Just like source invariance itself, its benefits also come in degrees. The more invariant the encoding is across a given set of sources of uncertainty and the more sources it is invariant across, the easier it will be to learn new tasks.
This kind of local adaptivity is particularly appealing because it makes plausible that source invariance could emerge from processes that learn in a series of small changes. This suggests that source invariance could have emerged as a consequence of evolution or learning in biological neural networks, and may emerge from gradient-descent learning in neural network models.

Source invariance is related to two broader challenges that arise in analysing representations.
The first is the general problem of invariant decoding. Just as a neural population encoding all relevant nuisance variables contains all information necessary to compute uncertainty, retinal responses to a given image contain all the information used by downstream regions in the ventral stream to compute more abstract information about this image such as category. So in principle, we could decode image category from retinal activations. However, whereas inferotemporal cortex (IT) explicitly encodes category information, the retina does not. To support this point empirically, we may argue that IT clusters stimuli based on their category and the retina does not. Alternatively, we may argue that we can read out category information from IT with a linear decoder, but would require a much more complicated readout model (something like the ventral visual stream) to decode category information from the retina.
As an important difference to this general problem, source invariance is less dichotomous. Whereas a representation likely represents either pixel-level information (as in the retina) or category-level information (as in IT), but not both, the same neural population could plausibly represent both uncertainty and the nuisance variables.

This complication also arises in a second important topic in representational analysis: disentanglement. Here, we know that a neural population encodes multiple variables (for example color and shape) and we are trying to determine whether it represents them in an entangled manner or not. These questions are often resolved using e.g.\ cross-decoding analyses, just like we are proposing for source invariance. However, disentanglement is usually concerned with variables that could, in principle, be represented independently of each other. Uncertainty, on the other hand, is independent from the relevant nuisance variables and any representation will have to trade off between representing one or the other.

Methods pertaining to both of these topics have been investigated in depth and are likely to be useful for analyzing source invariance, as well.
However, source invariance also poses new theoretical challenges, as it is concerned with representations of multiple, interdependent variables.

\subsection{A test for task-transferable uncertainty decoding}
\label{sec:task-transfer}

\begin{figure}
    \centering
    \includegraphics{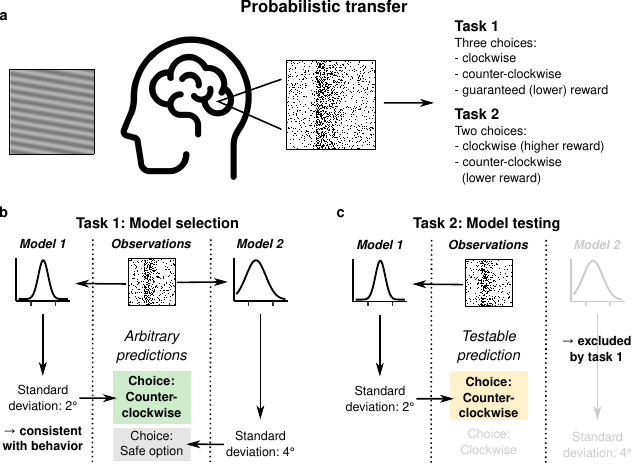}
    \caption{Probabilistic transfer. \textbf{a}, The same representation can be implicated in multiple tasks involving uncertainty (see Box~\ref{box:marginalization} for description of these tasks). \textbf{b}, On the first task, any behavior is consistent with some probabilistic model. This can serve as a model selection mechanism: if the representation were probabilistic, what would the generative model look like? \textbf{c}, While this generative model does not generate a testable prediction on the first task (as it was selected on the basis of its alignment with behavior on that task), it then generates a testable prediction on the second task.}
    \label{fig:prob-transfer}
\end{figure}

How can we test whether a representation of uncertainty is probabilistic when the underlying generative model is not defined? We have seen that the lack of constraints on the generative model poses a difficult problem for claims that a representation is used in a way that approximates the rules of probability: any computation can be cast as an instance of (approximate) probabilistic inference by assuming the right model. If our definition of probabilistic computation is so broad as to accommodate all computations, then there is nothing distinct about probabilistic computations and the answer to the question whether a representation is probabilistic is in the eye of the beholder: It depends only on whether we accept as plausible the generative model that enables us to interpret the computation as probabilistic inference.

The solution is to expand our focus to how subjects generalize across tasks (\cref{fig:prob-transfer}a). If a subject is using a generative model and a probabilistic computation, we can use their performance on the first task to characterize the generative model they could have used (\cref{fig:prob-transfer}b). We can then use the constraints the generative model imposes to predict their behavior on a second task (\cref{fig:prob-transfer}c). If an organism is using a probabilistic representation, probability theory provides a clear justification for these predictions. But if the subject is using a non-probabilistic representation, there is no reason to expect probability theory to make accurate predictions. We call this generalization ``probabilistic transfer." On the one hand, the subject transfers the generative model to the new task. On the other hand, we as researchers use our estimate of the subject's generative model to predict their behavior in the new task. This prediction provides a concrete and specific test of the hypothesis that the subject is using a probabilistic representation.

More concretely, we propose testing for two probabilistic computations (``distinctly probabilistic computations''): marginalization and change of variables. After describing them, we illustrate why some other probabilistic computations do not make testable predictions. In particular, we will describe why MAP estimation cannot provide a test for probabilistic representations even with probabilistic transfer.

\paragraph{Marginalization.}
The first distinctly probabilistic computation is marginalization. Marginalization is the process of computing the expectation of a function of a random variable: $\mathbb{E}_{x\sim p(x)}[f(x)]$. This notation means that the variable $x$ is distributed according to $p(x)$ and we take the expectation of $f(x)$ over this distribution. If $x$ is a discrete random variable, marginalization takes the shape $\sum_{i=1}^nP(x=x_i)f(x_i)$. If $x$ is a continuous random variable, marginalization takes the shape $\int f(x)p(x)dx$. Of course, expected values need not be computed as a sum or integral. To marginalize means to approximate the mapping from a probability distribution to the expected value. To give a simple example, consider $f(x)=x^2$ for a normally distributed variable with mean $\mu$ and standard deviation $\sigma$. In this case, the integral reduces to the expression $\mu^2+\sigma^2$ and marginalization would mean approximating this expression. (We say more about how the brain may approximate marginalization in \cref{sec:computations}.)

Our use of the term ``marginalization'' is broad. It captures a number of probabilistic computations that sometimes go by different names, but have the common feature of collapsing across one or more variables that are not of interest in their own right, but need to be considered to infer other variables that are of interest. Consider two particular uses of marginalization in this broad sense: 
1) Marginalization can be used to transform a joint probability distribution, such as $p(x,y)$, into a probability distribution over a subset of the random variables, such as $p(x)$, a computation usually known as ``marginalizing out'' variables that is relevant e.g. for hierarchical Bayesian inference \citep{charlton_environmental_2023}. 
2) Marginalization can also be used to calculate the expected value of a reward function. The probability of each possibility is multiplied by its utility, and the expected value is the sum of all those products. 
More generally, marginalization captures taking the expected value of any variable over a probability distribution. This estimate minimizes the expected squared error between that variable and our estimate. 
Since in each case we take an expectation over a probability distribution, we will use marginalization to refer to all of them.

Using probabilistic transfer, we can test whether a subject is marginalizing over a probability distribution (see \cref{box:marginalization}). To see this, consider a downstream computation that is learned under a deterministic probability distribution (i.e.\ a distribution whose mass is concentrated in one value). In that case, the expected value is simply $f(x)$, where $x$ is the value taken by the probability distributions. A subject who has experienced the outcome of this computation for this small subset of possible distributions and uses a probabilistic model knows the outcome the computation should take for all other probability distributions: namely $\mathbb{E}[f(x)]$. Suppose, for example, that we represent a normally distributed variable with mean $\mu$ and standard deviation $\sigma$. As explained above, if $f(x)=x^2$, we know that $E[f(x)]=\mu^2+\sigma^2$. Suppose we observed that a subject matches these predictions when generalizing the quadratic function to new levels of uncertainty. Under a nonprobabilistic account, there is no reason why the quadratic function should lead to this exact pattern of generalization. Under a probabilistic model, the reason is clear.

The more functions we observe probabilistic transfer for, the more implausible the nonprobabilistic account becomes. If $f(x)=\exp(x)$, its expected value would be $\mathbb{E}[f(x)]=\exp(\mu+\sigma^2/2)$. Because the observed behavior in each of these examples generalizes to new situations in a way that is consistent with the probabilistic computation of marginalization and is hard to explain otherwise, we consider it evidence in favor of a probabilistic computation. 

\begin{figure}
\begin{mdframed}[style=boxstyle, myboxedtitle={Box~\addtocounter{box}{-1}\refstepcounter{box}\thebox\label{box:marginalization}: Example experiments for testing marginalization}]
\paragraph{Learning a new utility function (Figure~\ref{fig:prob-transfer}).}
We might ask participants to look at a visual grating when there are multiple sources of uncertainty about the grating's orientation, such as low contrast or noise added to the pixel intensities. They are asked to decide whether an arrow appears rotated clockwise or counterclockwise of the grating's orientation. Subjects get rewarded for correct decisions. However, if they are unsure, they can opt out and receive a smaller reward with certainty. The way to maximize reward in this task is to compute the expected reward (probability times value) for choosing clockwise or counterclockwise, and pick the response with the larger expected reward of these two if it exceeds the opt-out reward (and opt out otherwise).

As we have seen, one can often construct a combination of a generative model and a utility function for which participants' behavior can be cast as marginalization (here computing the expected reward for each possible action). To demonstrate that this description actually captures the underlying inference, we can train participants on a second task with a different reward function. For example, we can remove the opt-out option and make rewards asymmetric, so that on each trial one choice, if correct, will yield a greater reward than the other choice, if it were correct. It is important that participants be exposed to the new task under a limited range of nuisance variables present in the original, so as to prevent them from learning the new mapping from stimuli to optimal behavior directly from experience. Once participants have learned the new utility function, we can introduce the full range of nuisance variables from the original experiment. If participants are able to take expectations of the new utility function in a way that our model of the first task captures, this would provide strong evidence for probabilistic computation as it requires them to generalize to new levels of uncertainty without any training. Moreover, if a neural population encodes the participant’s uncertainty about orientation in a manner invariant to the source of that uncertainty, this area represents a plausible first candidate as to the location of that probabilistic representation.

\paragraph{MAP estimation over a marginal distribution.}
Another instance of probabilistic transfer would be computing marginal probabilities from a joint distribution, a process often called “integrating out” uncertainty about a variable. As an example, suppose we have a probabilistic model of how participants represent the speed of a car and its distance from their own position (determining this model may involve observing the participants' behavior on a separate task). We can train participants to predict the most likely time at which this car will arrive at their position (the MAP estimate). This would involve computing arrival time as a function of speed and distance. There are multiple combinations of speed and distance that will lead to the same predicted arrival time. If a car is further away, but drives faster, it may arrive at the same time as a car that is closer but slower. To determine (and maximize) a probability distribution over arrival times, participants would have to marginalize over these different combinations. After being trained on this MAP estimation task under good viewing conditions, where their uncertainty should be low, we can test their behavior under bad viewing conditions, where their uncertainty should be high. If they use probabilistic representations and our model of their behavior is correct, it should be able to predict their behavior in this new context. Further, different factors can influence viewing conditions, for example whether it is day- or nighttime and whether it is foggy or not. If a neural population encodes uncertainty about speed and distance invariant to their source, measuring neural activity during task performance would, again, enable us to localize the probabilistic representation.
\end{mdframed}
\end{figure}

\paragraph{Change of variables.}
The second computation that can be tested using probabilistic transfer is the transformation of a probability distribution over one variable to a probability distribution over some function of that variable. Probability provides the optimal way to compute such transformations, which, if the function is invertible, is given by the change of variables formula\footnote{If the probability distribution is one-dimensional, for example, change of variables takes on the form $p_Y(y)=p_X(g(y)) \cdot |g'(y)|$, where $g$ is the inverse of $f$. There is an equivalent formula for multi-dimensional probability distributions. If the function is not invertible, computing the transformation will also involve a form of marginalization. More generally, we can derive a formula that, for any function $f(x)$, computes the probability distribution over the transformed variable. This generalizes  computing marginal probability distributions and change of variables and would allow us to formulate tests of probabilistic transfer as well. To highlight the difference between the two computations, we focus on invertible transformations here.}.
Suppose, for example, we learned to use the speed of a car to predict the time period it would take this car to come to a halt if its driver applied the brakes. The faster the car, the longer this would take. This function would therefore have a unique time period associated with each speed and thus be invertible. To generate a probability distribution over stopping times, we would use a change of variables transformation. 

Like marginalization, we can test for change of variables using probabilistic transfer (see \cref{box:change-of-variables}). To see how, consider a change of variables that is followed by computing the most likely value. A one-to-one transformation $f$ can be learned for a small subset of all possible probability distributions, for example the set of deterministic probability distributions. In that case, the MAP estimate of the transformed distribution would simply be $f(x)$. Having been learned on this subset, change of variables tells us how to apply this transformation to other probability distributions. For example, if $y=f(x)=\exp(x)$ and the represented probability distribution is a normal distribution with mean $\mu$ and standard deviation $\sigma$, then the MAP estimate over $y$ is given by $\exp(\mu-\sigma^2)$. Why the exponential function should lead to this exact transformation seems quite arbitrary without probability theory, but with probability theory, it is obvious.

\begin{figure}
\begin{mdframed}[style=boxstyle, myboxedtitle={Box~\addtocounter{box}{-1}\refstepcounter{box}\thebox\label{box:change-of-variables}: Example experiment for testing change of variables}]
In \cref{box:marginalization}, we considered a function that collapses  combinations of speed and distance that entail the same arrival time. Now imagine a variant where participants are asked to predict the travel time of the car between two intersections. Distance is now fixed across trials and participants are left to learn a monotonic function mapping speed to travel time (the faster the car, the shorter the travel time between the intersections). Whereas MAP estimation of arrival time involved computing a marginal distribution, MAP estimation of travel time now requires applying the change of variables formula. Again, participants could be trained  to estimate travel time under low-uncertainty conditions. Exposing them later to  high-uncertainty conditions, would reveal whether they use probabilistic representations. The hypothesis of probabilistic mental representations with change of variables, again, entails unique predictions that can be tested behaviorally. Further, we would be able to localize the probabilistic representation by looking for a neural population encoding uncertainty about speed in a source-invariant manner.
\end{mdframed}
\end{figure}

\paragraph{MAP estimation is not distinctly probabilistic.}
There are many probabilistic computations other than marginalization and change of variables. We already considered some of the most influential: MAP estimation and Bayes' rule. In \cref{sec:map}, we laid out how model indeterminacy prevents these computations from making testable predictions. We also established that any MAP estimate over a probability distribution has an equivalent interpretation as maximizing an objective function. Probabilistic transfer has allowed us to deal with model indeterminacy in the case of marginalization and change of variables. Can it do the same for MAP estimation and Bayes' rule? The answer is no. As we explain below, while a transfer experiment would allow us to make testable predictions, these predictions would still not be able to distinguish MAP estimation from a computation that maximizes an objective function.

To see why, we return to our example of estimating a car's speed from auditory and visual cues. Suppose we presented the subject with an auditory cue $a$ (for example the sound of a car). The reliability of this cue would vary from trial to trial. From a probabilistic perspective, we could determine the subject’s prior over car speed $p(s)$ as well as the conditional probability $p(a|s)$. For a single perceptual cue, the MAP estimate would then be given by maximizing $p(s)p(a|s)$. As we have already seen, finding that such a generative model could have explained the subject’s behavior as MAP estimation is not evidence for probabilistic representation. More specifically, there are two problems: first, any estimate of speed can be interpreted as the MAP estimate for some generative model; and second, any MAP estimation can be interpreted more parsimoniously as maximizing an objective function with a regularizer. In this case, maximizing $p(s)p(a|s)$ is equivalent to maximizing $\log p(s)+\log p(a|s)$. As we will see in the next paragraph, testing the generative model's predictions on a second task can help us address the first issue, but not the second one.

In particular, we can also determine the subject's responses to a visual cue $v$ (without exposing them to an auditory cue at the same time). This would allow us to determine $p(v|s)$ in the same way that we had determined $p(a|s)$. As a transfer task, we may now expose the subject to a visual and auditory cue at the same time. Probability theory would imply that if the visual and auditory cue are conditionally independent the subject should multiply the respective conditional probabilities and thus maximize $p(s)p(v|s)p(a|s)$. This would make testable predictions for the subject’s behavior on this second task. However, whereas for marginalization and change of variables it was unclear why a nonprobabilistic account would make the predictions that a probabilistic account had made, there are clear alternatives to the MAP account that do not rely on probability theory, yet make identical predictions. Namely, we could also simply assume that the subject had summed up the two individual objective functions and the regularizer, which would result in the exact same predictions. Even with predictions on a second task, MAP estimation is not discernible from maximizing an objective function\footnote{That is not to say that investigations of this type would be uninformative. Indeed, they may help us create more constrained tests for whether the brain estimates latent variables by maximizing objective functions. However, they do not tell us whether the brain uses probabilistic representations.}.
We therefore face the same situation as in \cref{sec:map}: if we define a MAP estimate to be a probabilistic representation, then any representation maximizing an objective function would also have to be considered probabilistic.

This is not to suggest that MAP estimation is not a probabilistic computation; it clearly is. MAP estimation simply cannot be used to \emph{test} whether a given representation is probabilistic.
We refer to computations that make such tests possible as \emph{discernible probabilistic computations}. Other, non-discernible probabilistic computations (such as MAP estimation, incorporating a prior, or cue combination) can still be crucial for successful use of a probabilistic representation. They simply cannot be used to test whether the representation is indeed probabilistic.

The feasibility of testing for a probabilistic representation also depends on which probability distributions can be represented and which downstream readouts can be used. For example, if a probabilistic representation of a normal distribution can only use linear readouts, marginalization would make the same predictions as taking the MAP estimate and then applying the linear function to that estimate. To distinguish a probabilistic representation from a heuristic representation of uncertainty, we would have to consider additional downstream computations, for example an asymmetric reward functions. This clarifies ongoing debates on whether or not normal distributions should be excluded from a definition of probabilistic representations (\cref{sec:ideal-observers}).

Are marginalization and change of variables the only discernible probabilistic computations? This question is difficult to answer. There may be additional discernible probabilistic computations. Take, for instance, sampling. Mere stochasticity should not be taken as evidence of probabilistic representations. More complicated sampling processes that approximate a posterior, however, may have discernible signatures. In this case, the principle of probabilistic transfer makes concrete how one could test such signatures without being subject to model indeterminacy.

In sum, we define probabilistic representations as representations of uncertainty with downstream computations that are discernibly probabilistic. This definition is testable, as the downstream computations on different tasks must be consistent with the same probability distribution. Thus, whereas on a single task, heuristic and probabilistic representations of uncertainty can be equally consistent with a subject's behavior, observing the subject's behavior across many different tasks allow us to break the tie. In particular, as we observe behavior that is consistent with probabilistic computations over the same probability distribution across more and more tasks, we can exclude more and more heuristic representations. In turn, observing behaviors that are inconsistent with a shared probability distribution (or at least require a more and more complicated distribution) would exclude more and more probabilistic representations --- and suggest that the representation is heuristic.

\paragraph{Properties of probabilistic transfer.}
We are now in a better position to appreciate why we needed to exclude representations with a fixed level of uncertainty: Probabilistic transfer allows us to test probabilistic representations by generating predictions for levels of uncertainty that our participant had not previously experienced in the context of the task. This is not possible if the represented probability distribution does not vary in its uncertainty. Could we instead generate such predictions for, say, locations of the probability distribution that the participant had not previously experienced? No. To see this, let us first consider deterministic probability distributions. These distributions clearly do not represent uncertainty and only vary in their location, that is the value $x$ in which all probability mass is concentrated. Marginalization over a function $f$, will simply lead to the outcome $f(x)$ and change of variables will simply transform this distribution to a probability distribution with all probability mass concentrated in the value $f(x)$. So generalizing to new locations of this probability distribution is impossible without any constraints on the underlying function $f$.
More generally, probability distributions that do not represent uncertainty make no distinct predictions for marginalization or change of variables. Consider, for example, for a normal distribution with a varying mean $\mu$ (e.g. the speed of the car) and a fixed standard deviation (representing our uncertainty about the speed). If marginalization made testable predictions for this set of distributions, we would have to be able to observe responses to a limited set of values $\mu$ (speeds) and use this to predict responses to all other values of $\mu$. Such out-of-distribution predictions would require that marginalization does not allow for arbitrary input-output relationships between $\mu$ (speeds) and responses $g(\mu)$ (for example a behavioral response that involves marginalization). But we can interpret any function $g(\mu)$ as the result of such a marginalization over an underlying function $f(x)$. More formally, for any function $g$, we can find a function $f$ that allows us to write $g(\mu)=\mathbb{E}_{x\sim\mathcal{N}(\mu,\sigma^2)}[f(x)]$. So the knowledge that a downstream computation should be the result of marginalization or change of variables, in this case, would not constrain the possible functions $g$ at all, and therefore make no testable predictions.

Probabilistic transfer, like source invariance, comes in degrees. To be considered probabilistic, a representation of uncertainty does not need to be capable of marginalization and change of variables for all possible downstream computations, and it does not need to approximate the laws of probability theory perfectly. Rather, the more functions the representation can use in discernible probabilistic computations, and the better it can approximate these computations for these functions, the more compelling it is to consider the representation probabilistic.

Probabilistic transfer, like source invariance, has benefits for the animal. If the animal had a heuristic representation of uncertainty, it would have to learn a new task dependent on this representation for each level of uncertainty. With a probabilistic representation, it can learn this task for a limited range of uncertainties, and then generalize to new levels of uncertainty, resulting in a more efficient learning process. This advantage also comes in degrees: if the animal is better at approximating probabilistic computations and can do so for more functions, its learning will be more efficient. So probabilistic transfer, just like source invariance, could plausibly emerge from a learning rule that uses small adjustments driven by task performance (such as gradient descent and most models of biological learning and evolution).

\paragraph{Probabilistic transfer and Bayesian transfer.}
We are not the first to suggest using transfer in the context of probabilistic computations. \citet{koblinger_representations_2021} lay out how probability distributions can be flexibly used across multiple contexts, though they focus on usefulness more than testability. Relatedly, probabilistic transfer overlaps with a framework called Bayesian transfer \citep{trommershauser_statistical_2003,trommershauser_optimal_2005,maloney_bayesian_2009,kiryakova_bayesian_2020}. Bayesian transfer is an application of Bayesian Decision Theory to perceptual decision making. Within Bayesian transfer, a task is defined by a likelihood function, a prior distribution, and a utility function. Using these three elements, a participant could compute a posterior distribution and the expected utility over the posterior, and maximize this utility. On a single task, however, the same result could be achieved by a generic learning algorithm. More importantly, as we have pointed out here, almost any behavior can be explained by applying this framework with a suitable combination of generative model and utility function. Bayesian transfer attempts to avoid these pitfalls by introducing a second task, in which either the likelihood, prior, or utility function is replaced. A generic learning algorithm would have to learn the task from scratch, associating different responses in each stimulus condition to different outcome values, whereas an algorithm relying on Bayesian Decision Theory would be able to re-use its previously acquired knowledge about the task structure.

Many cases of Bayesian transfer are consistent with the notion of probabilistic transfer, and vice versa. While they overlap, however, probabilistic transfer is more specifically focused on distinguishing probabilistic from non-probabilistic representations. Bayesian transfer is limited in two ways in this regard. First, in some Bayesian transfer settings, the combination of generative model and utility function makes behavior indistinguishable from MAP estimation and thus indistinguishable from non-probabilistic explanations \citep{sato_how_2014}.
Second, some behaviors which do not have clear non-probabilistic explanations fall outside of the scope of Bayesian transfer. As we saw above, transforming an uncertainty representation via a change of variables computation would be difficult to explain without invoking probability, but it does not necessarily involve Bayesian Decision Theory and thus is not an example of Bayesian transfer. The same is true of integrating out a variable from a joint distribution. Probabilistic transfer is focused on generalizing distinctly probabilistic computations across task settings. Thus, some examples of Bayesian transfer are not probabilistic transfer, and some examples of probabilistic transfer, such as change of variables or computation of marginal probabilities, are not Bayesian transfer. Further, Bayesian transfer is mostly concerned with clarifying the benefits of probabilistic representations, and therefore more comfortable with assuming an ``optimal'' generative model than we are. In spite of these differences, Bayesian transfer, our framework, and the perspective by \citet{koblinger_representations_2021} all emphasize the importance of probability theory to task-flexible representations of uncertainty. Bayesian transfer, in particular, has resulted in the best evidence so far for the existence of probabilistic representations.

\section{Evidence, Challenges, and the Way Forward}
\label{sec:evidence}

There are a lot of open questions about probabilistic representations.
In this section we will describe how future research grounded in source invariance and probabilistic transfer might answer those questions, focusing on probabilistic representations in perception.
\cref{sec:which-systems} will focus on the question of which systems have probabilistic representations.
\cref{sec:computations} will discuss the algorithms potentially underlying their use.
Finally, in \cref{sec:image-computable}, we will describe how our theory of probabilistic representation might help us build better image-computable models, a goal at the intersection of computational neuroscience and artificial intelligence research.

\subsection{Which systems have probabilistic representations?}
\label{sec:which-systems}
There is strong evidence that sensory regions are able to respond to individual nuisance variables and that subjects can account for them in a manner that allows for near optimal performance on individual tasks \citep{ma_bayesian_2006,van_bergen_sensory_2015,walker_neural_2020}.
While this evidence is consistent with probabilistic representations, we can explain it without invoking them, as well.
First, it is consistent with sensory representations of nuisance variables, which then influence post-perceptual representations of uncertainty.
Second, it is consistent with representations of uncertainty that do not invoke probability.
As we reviewed in \cref{sec:our-theory}, studies would need to demonstrate source invariance and probabilistic task transfer to provide evidence for representations of uncertainty that are uniquely probabilistic.

One recent experiment has demonstrated source invariance in a representation of uncertainty in visual cortex \citep{henaff_representation_2020}.
This study found multiple sources of noise were encoded in the gain variability of individual neurons in macaque V1, but it did not include any behavioral results, and so there was no way to characterize whether and, if so, how the uncertainty was used. Another study has found that the brain may represent internal sources of uncertainty (e.g. uncertainty due to neural noise) with higher accuracy than external sources of uncertainty (e.g. stochastic environments) \citep{aston_different_2023}.

Other studies of uncertainty representation have also recognized the challenges arising from investigating a representation of subjective belief (rather than an objective property) \citep{walker_studying_2022}. \citet{bounmy_characterization_2023} studied representations of uncertainty about subjects' estimates of how often different Gabor patches are presented to them. Such tasks that do not probe perceptual uncertainty directly circumvent the issue of representational indeterminacy (as there are no nuisance variables). Other studies attempt to circumvent the issue of representational indeterminacy by holding the nuisance variables constant \citep{van_bergen_sensory_2015,walker_neural_2020}. They suggest that any remaining differences in uncertainty across trials must arise from internally generated uncertainty. As the nuisance variable is held constant, this approach may circumvent the issue of representational indeterminacy.
As we note in \cref{sec:uncertainty}, it is not clear to us whether this truly avoids representational indeterminacy or could still confound a representation of uncertainty with a representation of estimated nuisance variables (which may vary even when the nuisance variable is held constant). We therefore suggest that a direct test of whether a given representation is invariant to different nuisance variables provides important complementary evidence.

Other studies indicate that some probabilistic representation is used for marginalization in perceptual decision making.
\citet{whiteley_implicit_2008} asked human subjects to judge which of two visual stimuli was shifted further to the left or right.
Because different rewards were associated with each option, an optimal decision required subjects to weigh the options by their respective probability.
Despite not being provided immediate feedback (as would be required if they learned the task from scratch), the subjects were able to perform almost optimally.
This is consistent with the use of probabilistic representations of both grating locations.

Importantly, existing studies testing Bayesian transfer often do not address the issue of model indeterminacy, instead assuming that humans use an ideal observer model and testing whether they can transfer that generative model. This is in line with a view emphasizing the usefulness of Bayesian transfer \citep{koblinger_representations_2021}. It also provides a solution for the problem of inferring a generative model from behavior alone. However, as we note in \cref{sec:ideal-observers}, the concept of an optimal generative model might not always be well-defined outside of specific behavioral tasks. Further, a test of Bayesian transfer that assumes an optimal model may miss probabilistic representations that implement a different generative model.

Finally, as we detailed above, there are notable differences between Bayesian transfer and probabilistic transfer. A study of Bayesian transfer in the visual system by \cite{sato_how_2014} illustrates why these differences matter for detecting probabilistic representations. This study tested whether subjects could integrate a learned prior and likelihood function that they had not previously experienced together, in order to maximize a utility function --- an instance of Bayesian transfer. In doing so, it used a normal prior and likelihood combined with a utility function of negative squared error. Because minimizing squared error for a normal posterior is equivalent to computing the mode, Bayes-optimal behavior in this setting is equivalent to MAP estimation, which can always be interpreted in terms of a non-probabilistic objective function. Because there is a non-probabilistic explanation for the generalization behavior, we do not consider this an instance of probabilistic transfer.

\subsection{What are the underlying computations?}
\label{sec:computations}

So far we have been concerned with showing that a representation is probabilistic regardless of format.
While probabilistic transfer involves characterizing a probability distribution representing uncertainty, it makes no assumptions about how the brain encodes that uncertainty, what mechanisms underlie its use, or what distributions it can encode.
To more fully understand probabilistic representations, we need to characterize the computations that produce and make use of them.
A variety of computational models have been proposed to explain how sensory cortex might represent probability distributions \citep{knill_bayesian_2004,jazayeri_optimal_2006,pouget_probabilistic_2013}.
These coding schemes could involve single neurons or populations, which can express probability distributions as parameters, samples, or explicit representations of probability or log probability \citep{rao_bayesian_2004,jazayeri_optimal_2006,yang_probabilistic_2007}.
Different dimensions of their activity could be used to represent uncertainty: vectors of population activity, spatial or temporal variability, and variability in neural gain have all been proposed as vehicles of probabilistic representation.
In this section, we will outline some popular probabilistic models in computational neuroscience, and examine them in light of source invariance and probabilistic transfer.

There are two reasons to review these models in terms of the properties we have proposed here.
First, if the brain makes use of probabilistic representations, we would like to understand how.
What mechanisms make this possible?
To understand what computations could support probabilistic representation, computational models need to exhibit source invariance and probabilistic transfer.
Second, the computational models we will describe here face the same problems of representational and model indeterminacy we outlined above.
If they do not exhibit source invariance and transfer, there will always be non-probabilistic models which could account for the same observations. Our theory is therefore also relevant for testing models of how probability distributions are encoded.

One influential encoding model of probability distributions has been probabilistic population codes \citep[PPCs;][]{zemel_probabilistic_1998,ma_bayesian_2006}.
PPCs assume that the activity of a group of neurons can be described as following an exponential family likelihood function whose parameters depend on a certain stimulus.
This activity allows for the linear decoding of the parameters of a posterior distribution over this stimulus.
As discussed before, the ability to decode probability distributions from neural activity is not sufficient evidence for claims that the brain represents probability distributions.
However, PPCs make a more specific claim: the variability of sensory neurons is such that the parameters of the optimal posterior distribution can be decoded in a neurally plausible manner.
For example, the posterior mean and variance for orientation can be linearly decoded from a population of independent Poisson neurons with overlapping receptive fields tuned to the different orientations.
In the most general form of PPCs, the population activity depends on a set of nuisance parameters.
The uncertainty resulting from these nuisance parameters, according to the theory, would be encoded in the gain, i.e.\ the overall amplitude of each unit’s activity.
More specifically, the theory proposed that higher levels of uncertainty lead to lower levels of gain.

There has been no research on whether PPCs can allow for the types of generalization necessary for probabilistic task transfer.
There has been some theoretical work exploring classes of computations that could implement marginalization for a wide range of tasks.
In particular, \citet{beck_marginalization_2011} demonstrate that a network with a quadratic nonlinearity and divisive normalization could accommodate marginalizing over several functions.
However, while networks with this architecture can learn to approximate marginalization, probabilistic transfer relies on downstream computations that are \emph{constrained} to implementing marginalization and PPCs, in their current form, can therefore not help us explain probabilistic transfer.
To investigate how PPCs could implement probabilistic transfer, we would have to find an architecture implementing marginalization or change of variables in a way that can be transferred to multiple tasks.
Such architectures could be constructed from first principles or by meta-learning processes \citep{wang_meta-learning_2021}.

PPCs represent parameters of probability distributions in the (temporally averaged) activity of a population of neurons.
Another approach to modeling probabilistic neural representations assumes that neural variability (usually changes in firing rate across time) represents samples from a posterior distribution given sensory inputs \citep{moreno-bote_bayesian_2011,sanborn_sampling_2017}.
Sampling models are usually constructed such that populations of neurons represent posteriors over latent variables in an internal generative model.
One common example in vision is the Gaussian scale mixture model \citep{wainwright_scale_2000,orban_neural_2016}, which assumes that visual sensory input is roughly modeled as a linear combination of image features, multiplied by a global contrast value.
Under this generative model, neurons in visual cortex can be modeled as sampling from the posterior distribution of these image features.
Notably, sampling models and PPCs are not necessarily distinct from each other. \citet{shivkumar_probabilistic_2018} demonstrate that a sampling-based representation under a linear Gaussian generative model, when presented with oriented gratings, can also be cast as a probabilistic population code over orientation \citep[see also][]{lange_bayesian_2020}.

One strength of sampling-based representations is that marginalization is easy to implement.
Individual neurons in sampling codes implicitly represent marginal posteriors as they represent one dimension of the represented information. Further, the expectation of any function can be evaluated by averaging samples over time.
The same is true of change of variables.
Any transformation of the represented samples will represent samples from the transformed posterior.
However, to our knowledge there has been no work on sampling models showing that this marginalization or change of variables takes place, either through behavior or downstream effects on other neurons.

Another promising computational model of probabilistic representations are Distributed Distributional Codes \citep[DDCs;][]{zemel_probabilistic_1998,zemel_distributional_1999,sahani_doubly_2003}.
DDCs represent random variables by taking the expected value of different functions $T_i(x)$ of that variable.
This set of functions is called the basis set.
It has no constraints other than that it should be sufficiently versatile as to accurately specify the encoded probability distribution. \citet{vertes_flexible_2018}, for example, use random linear-nonlinear functions with a sigmoidal nonlinearity, while \citet{zemel_probabilistic_1998} use Gaussian kernels.

Because they are centered around expected values of downstream functions, DDCs naturally support probabilistic transfer.
More specifically, \citet{vertes_flexible_2018} propose to approximate marginalization over a function $f$ by approximating this function as a linear combination of the basis functions.
The expected value of this function is then approximated by the same linear combination of the expected value of the basis functions.
The more basis functions we have, the better we approximate $f$ and, thus, the rules of probability theory.
We can implement change of variables in the same way.
After all, a DDC would encode a probability distribution that is transformed by $f$ simply by encoding the expected values $\mathbb{E}[T_i(f(x))]$.
So again, we can approximate change of variables by approximating $T_i(f(x))$ as a linear combination of basis functions $T_i(x)$.
DDCs therefore not only support a plausible implementation of probabilistic transfer, but also make concrete predictions for the quality of this transfer as dependent on whether we can approximate the downstream function as a linear combination of basis functions $T_i(x)$.

Besides designing algorithms that can implement probabilistic transfer (or testing existing algorithms), we could also investigate how the brain encodes uncertainty by exploring what kinds of computations emerge from different types of selection pressures in a generic computational mechanism such as a connectionist network \citep{orhan_efficient_2017,echeveste_cortical-like_2020,tyulmankov_meta-learning_2022}.
We have already described one paper \citep{beck_marginalization_2011} demonstrating that neural networks can learn to perform tasks which require marginalization.
Moving forward in this direction would involve asking under which circumstances and with which inductive biases such networks might be able to learn to perform probabilistic transfer.

\subsection{Building image-computable probabilistic models}
\label{sec:image-computable}

One issue with many computational neuroscience models is their lack of scalability.
In recent years, there has been increasing acknowledgment of the virtuous cycle between neuroscience, which provides inspiration for computational mechanisms, and machine learning/artificial intelligence, which engineers scalable real-world solutions.
From the perspective of psychology and neuroscience, we would like our models, like those in machine learning, to have inputs and outputs that are similar to those of biological systems \citep{macpherson_natural_2021}.
From the perspective of machine learning, we would like to understand how brains perform complex tasks such as object recognition or natural language processing to guide the engineering of new learning systems \citep{hassabis_neuroscience-inspired_2017}.
Probabilistic representations of uncertainty are a central topic for machine learning and statistics --- indeed, research in statistics and machine learning has given rise to many of the key methodologies that neuroscience suggests may explain how the brain represents probability, from basic estimation of probability distributions to variational inference.

At first glance, machine learning may appear to not suffer from the indeterminacy issues discussed in this article. After all, if we define a specific architecture in probabilistic terms, we can be confident that it contains a probabilistic representation. Consider, for example, a deep neural network outputting a probability distribution over predictions of a certain target variable. We can use this variable in probabilistic terms in a downstream computation (for instance another network). We can also assess whether the probability distribution is well-calibrated, i.e.\ provides a useful estimate of the overall error rate \citep{guo_calibration_2017}.

However, the brain may represent uncertainty about many latent variables.
Analysing latent probabilistic representations in models such as deep neural networks presents additional challenges.
Even if we provide the model with the infrastructure to represent probabilistic variables, it may use this infrastructure for some other computational purpose. This is arguably another instance of model indeterminacy: even if we explicitly define something as a probability distribution, it could be used as a computation that should not be interpreted as probabilistic (just as non-probabilistic computations almost always have a probabilistic interpretation).

For example, consider one of the most prominent examples of latent variable models in neural networks: the variational auto-encoder \citep{kingma_auto-encoding_2014,rezende_stochastic_2014}.
Given a particular input, this model represents a posterior distribution over a latent variable --- by all accounts, this is a probabilistic representation.
However, the probabilistic machinery for this latent distribution may simply serve as a regularization mechanism and replacing it by alternative regularization schemes can yield comparable performance \citep{ghosh_variational_2020}.
As a result, such a representation is arguably non-probabilistic according to our framework even though it is explicitly defined as a probability distribution.
Notably, current state-of-the-art variational auto-encoders have also done away with latent probabilistic variables entirely \citep{razavi_generating_2019}, illustrating that they do not play an important role in these models.

Of course, the reverse could also be true: we could design a neural network without any probabilistic machinery and its training process may give rise to an implicit probabilistic representation.
Methods from neuroscience can inspire the identification of such representations. In particular, \citet{van_bergen_sensory_2015} provide a method for finding a potential latent representation of uncertainty.
Our framework, on the other hand, provides a way of testing whether a given representation is probabilistic (i.e.\ whether this representation has explicitly probabilistic machinery associated with it or not).
Notably, multi-task learning has become an increasingly influential technology in modern machine learning \citep{vafaeikia_brief_2020,du_survey_2022}.
In particular, it is common practice to pretrain models on large amounts of data drawn from a general-purpose task before finetuning them on more specific use cases \citep{bommasani_opportunities_2022}.
Understanding the basis of generalization for these models has become increasingly important. Our framework for probabilistic representations provides one suggested mechanism by which models could improve their generalization from being trained on multiple tasks.

\section{Conclusion}
Researchers have long debated whether humans and animals have probabilistic representations.
A central issue in this debate has been a lack of agreement on what it means to have a probabilistic representation.
Here we have argued that this question of how a probabilistic representation is defined should be given careful consideration and have highlighted issues with prevalent frameworks.
We have then proposed forward our own theory, drawing substantially on existing work \citep{maloney_bayesian_2009,koblinger_representations_2021}.
Our theory is grounded in the principles of probability theory, renders the definition of probabilistic representations testable, and can help design experiments investigating probabilistic representations and their computational implementation.

Other researchers might find different definitions of probabilistic representation more useful.
Our article situates such definitions in a broader framework and highlights the roles different definitions can (and cannot) play depending on researchers' considerations.
We hope that this can ground a debate on when these different definitions are useful, which in turn may inspire new empirical investigations.
Regardless of the definition a particular investigation may rely on, we suggest that being clear about this definition is essential to making progress on understanding probabilistic representations in the brain.

\section*{Funding statement}
The authors acknowledge the support of NSF award 2240859, Andrew W. Mellon Foundation grant 1710-04974, the Summer Seminars in Neuroscience and Philosophy (SSNAP), the Presidential Scholars in Society and Neuroscience program at Columbia University, NSF 1707398 (Neuronex), and Gatsby Charitable Foundation GAT3708.

\small

\bibliography{references.bib}

\end{document}